\documentclass[12pt,twoside]{article}
\usepackage{a4}
\usepackage{latexsym} \usepackage{epsf}
\usepackage{cite}
\usepackage{amsfonts}

\begin{document}
\title{Spectral branes}
\author{D.V. Vassilevich\thanks{On leave from V.A. Fock
Department of Theoretical Physics,
     St. Petersburg University,
     198904 St. Petersburg, Russia.
     e-mail: Dmitri.Vassilevich@itp.uni-leipzig.de} \\
{\it University of Leipzig, Institute for Theoretical Physics}\\
{\it  Augustusplatz 10/11, 04109 Leipzig, Germany}}

\maketitle

\begin{abstract}
We study the objects (called spectral branes or $S$-branes)
which are obtained by imposing non-local spectral boundary
conditions at the boundary of the world sheet of the bosonic
string. They possess many nice properties
which make them an ideal test ground for the string theory methods.
Depending on a particular choice of the boundary
operator $S$-branes may be commutative or non-commutative.
We demonstrate that projection of the $B$-field 
on the brane directions (i.e. on the components
which actually influence the boundary conditions) is done with the help of the
chirality operator.
We show that the $T$-duality transformation maps an
$S$-brane to another $S$-brane. At the expense of introducing
non-local interactions in the bulk we construct also
a duality transformation between $S$-branes and $D$-branes
or open strings.
\end{abstract}

\section{Introduction}
Open strings with Dirichlet conditions at the boundary
of the world sheet \cite{Dai:1989ua,Leigh:1989jq} were
first considered as a rather exotic object. However,
a few year later \cite{Polchinski:1995mt} these objects
(the $D$-branes) became a center of a very important
development in the string theory. Therefore, it seems
natural to attempt to introduce different boundary
conditions in string theory and to study some properties
of the resulting objects.

In the present paper we suggest to impose a non-local
 spectral boundary conditions on coordinates of the endpoints
of open bosonic string. Such boundary conditions we introduced
by Atiyah, Patodi and Singer \cite{Atiyah:1980jh} to study
the Index Theorem. Since that time they appeared in
various mathematical and physical contexts (see e.g.
\cite{Grubb95,Dowker:2000sy,Mishchenko:1992vm,Falomir:1998as,Esposito:1997mw}
and references therein). Strictly speaking, spectral boundary
conditions were suggested for fermion theories. Here
we extend the scheme a little bit to use it for the bosonic
strings. To this end, we represent the Laplace operator
which appears in the equations of motion through the Dirac
operators. In the simplest case this may be done explicitly.
For more complicated background field we argue that such
representation is possible.

We call the object which is obtained by imposing spectral
boundary conditions on the boundary of the world sheet 
the spectral brane (or $S$-brane, for short). It
looks like a non-local mixture of the $D$-brane and the
open strings states. Namely, if we define a ``boundary helicity
operator'' (see eq. (\ref{helop}), the positive helicity
states are taken from $D$-brane, while the negative helicity
ones -- from the open string. The $S$-brane
shares some nice properties with more common string
models. For example, the string coordinates may be non-commutative
for a suitable choice of the boundary operator. As for the
$D$-brane, only part of the components of the $B$-field
enters actually the boundary conditions. Projection on
the ``brane directions'' is local (even though the boundary
conditions are non-local) and is done with the help
of a chirality matrix $\gamma_5$ which appears naturally
in the construction. Different parity properties of the $\gamma_5$
suggest that the low-energy effective theory may differ 
substantially from that for $D$-branes.

Duality transformations play a central role in string
theory. They relate strong and weak coupling limits of
various string models.
It is known that the $T$-duality interchanges
Dirichlet and Neumann boundary conditions. Since the $S$-brane is
a non-local mixture of the Dirichlet and Neumann states,
under $T$-duality it is transformed to another $S$-brane.
This means that $S$-branes may occupy a special place among
other string/brane models. It is also interesting that
by moving non-locality to the bulk action we may even
define duality transformation between $S$-branes and
$D$-branes or open strings.

The paper is organised as follows. In next section we consider
some consistency requrements which should be satisfied by
the boundary conditions and formulate spectral boundary
conditions for the bosonic string. We also find a local
projector of the $B$-field on $S$-brane and discuss non-commutativity
of the string coordinates. Sec.~3 is devoted to $T$-duality.
We construct in a standard way the duality transformation
which maps $S$-branes to $S$-branes and comment on $S$-brane --
$D$-brane and $S$-brane -- open string duality. Sec.~4 contains
discussion of the results and perspectives. Some mathematical
statements on the relations between Dirac and Laplace operators are 
collected in the Appendix. 

Throughout this paper we consider the simplest possible
geometry of the string. We suppose that the target
space metric is constant. In sec.~2 we also add a constant
$B$-field.

\section{Boundary conditions}\label{secbc}
Consider the action for the open bosonic string:
\begin{equation}
S=\frac 12 \int_M d^2x \sqrt{h} h^{ab}G_{\mu\nu}\partial_a X^\mu
\partial_b X^\nu \,. \label{opact}
\end{equation}
The string tension $\alpha'$ has been absorbed in $G_{\mu\nu}$.
For simplicity, we suppose that the target space metric $G_{\mu\nu}$ is
constant and the world-sheet metric $h$ is flat. 
We also assume Euclidean signature on the world
sheet and on the target space. Variation of the action
(\ref{opact}) with respect to $X$ gives:
\begin{equation}
\delta S= \int_M d^2x \sqrt{h} G_{\mu\nu} (\delta X^\mu)
(-\nabla_a \nabla^a )X^\nu -\int_{\partial M} d\tau 
(\delta X^\mu )\partial_\sigma X^\nu G_{\mu\nu} \,,
\label{varopa}
\end{equation}
where $\tau$ is the proper distance along the boundary.
$\partial_\sigma$ denotes partial derivative with respect to
an inward pointing unit normal. The first (volume) term in
(\ref{varopa}) generates the equations of motion, while
the second term should vanish due to the boundary conditions.

There are two consistent choices of local boundary conditions.
These are Dirichlet ($\delta X|_{\partial M}=0$) or
Neumann ($\partial_\sigma X|_{\partial M}=0$) conditions.
One can also impose different conditions on different components
of $X^\mu$ thus constructing various $D$-branes.

It we allow the boundary conditions to be non-local our choice becomes
much wider. On and near the boundary one can decompose $X$
into a sum of orthogonal functions
\begin{equation}
X^\mu (\sigma ,\tau) =\sum\limits_n \chi^\mu_n (\tau ) c_n(\sigma )\,,
\quad
\int\limits_{\partial M} d\tau \chi^\mu_n \chi^\nu_m G_{\mu\nu} =\delta_{nm}
\,,\label{ortsum}
\end{equation}
so that the surface integral in (\ref{varopa}) becomes a sum over
$n$ of a product of the boundary values of the Fourier coefficients $c_n$
and their normal derivatives:
\begin{equation}
\sum\limits_n (\delta c_n(0)) \partial_\sigma c_n (0)\,.
\label{bsum}
\end{equation}
$\sigma =0$ is the boundary.
To cancel this boundary term one should require that 
for any $n$ either $\delta c_n$ or $\partial_\sigma c_n $
vanishes at the boundary. This means that one may impose Dirichlet
or Neumann boundary conditions on different harmonics independently.

In this way one can construct an infinite variety of boundary
conditions. Most of them are too exotic to be physically
relevant. We restrict ourselves to the spectral boundary
conditions introduced by Atiyah, Patodi and Singer \cite{Atiyah:1980jh}
in their study of the index theorem for manifolds with boundary.
These boundary conditions are defined by an operator of Dirac
type associated with the Laplacian on $M$. In order to define this operator
let us introduce a moving frame $E_\mu^A$ on the target space
and let $X^A=E_\mu^A X^\mu$. The equations of motion now become
\begin{equation}
\Delta X^A=0,\qquad \Delta :=-\nabla^a\nabla_a \,.
\label{eom}
\end{equation}
Suppose that the target space is even dimensional. We introduce
$\gamma$-matrices such that 
\begin{equation}
\left(\gamma_a\gamma_b +\gamma_b\gamma_a\right)^{AB}=-\delta^{AB}h_{ab}
\,. \label{gammas}
\end{equation}
For more than two target space dimensions any representation of
the Clifford algebra (\ref{gammas}) is reducible. Effectively,
one should decompose the target space into a direct sum of 
two-dimensional subspaces and define standard $\gamma$-matrices
on each subspace. Then
\begin{equation}
\Delta =P^2\,,\qquad P=\gamma^a\nabla_a \,. 
\label{DP}
\end{equation}
Let us define a first order differential operator on 
the boundary $\partial M$
\begin{equation}
{\cal B}:=\gamma_\sigma \gamma^\tau \nabla_\tau \label{helop}
\end{equation}
which we call boundary helicity operator. This operator is
self-adjoint and, therefore, has a real spectrum. Let 
$\Pi_+$, $\Pi_0$ and $\Pi_-$ denote projectors on the
spaces with positive, zero and negative eigenvalues of
${\cal B}$ respectively. Spectral boundary conditions
are defined as:
\begin{equation}
\delta (\Pi X)|_{\partial M}=0,\qquad 
\partial_\sigma (1-\Pi )X |_{\partial M}=0 \,.\label{isbc}
\end{equation}
We will consider two cases: $\Pi =\Pi_+$ and $\Pi =\Pi_++\Pi_0$.

There is an inhomogeneous term in the Dirichlet
part of the boundary conditions (\ref{isbc}) $\Pi (X-X^{(0)})|_{\partial M}=0$
($X^{(0)}$ is fixed, $\delta X^{(0)}=0$). Due to the presence of
this term solutions of (\ref{isbc}) do not form a linear space.
To discuss hermiticity and Green functions we should
replace this problem by an associated linear problem. This can be
done by using the background field formalism (see below).
In the present context we can simply use translational
invariance to put $ X^{(0)}=0$. We arrive at linear homogeneous
boundary conditions:
\begin{equation}
(\Pi X)|_{\partial M}=0,\qquad 
\partial_\sigma (1-\Pi )X |_{\partial M}=0 \,.\label{sbc}
\end{equation}
Note, that with the conditions (\ref{sbc}) $\Delta$
is formally self-adjoint.\footnote{In spectral theory of the Dirac
operator instead of the second boundary
condition in (\ref{sbc}) it is common to require $\Pi P X|_{\partial M}=0$.
This guarantees existence of the spectrum of $P$. In our construction
the Dirac operator plays an auxiliary role. It is more important that
the variation problem (\ref{varopa}) is well defined. For this reason
we have to choose the second boundary condition as it is written
in (\ref{sbc}).}

Requiring that  coordinates of the endpoints of a 
string satisfy the conditions (\ref{sbc}) we
obtain a new object which we call {\it spectral brane}.
It is constructed from positive helicity states of
the $D$-brane system and negative helicity states
of the open string. The boundary conditions (\ref{sbc})
are non-local. This suggests that we are dealing
with certain collective excitations of fundamental
strings.

Let us switch on the $B$ field by adding to the
action (\ref{opact}) the following term
\begin{equation}
S_B=\frac i2 \int_M d^2x \epsilon^{ab}B_{\mu\nu}
\partial_a X^\mu \partial_b X^\nu \,.\label{Bvol}
\end{equation}
Again, we suppose for simplicity that the $B$ field is constant.
Then (\ref{Bvol}) contributes to the boundary term only in variation
of the string action. Total boundary term in this variation now
reads:
\begin{equation}
-\int_{\partial M} d\tau (\delta X^\mu)(G_{\mu\nu}\partial_\sigma
+iB_{\mu\nu}\partial_\tau )X^\nu \,.\label{Bbouvar} 
\end{equation}
Since the volume term in the variation is not changed, there is no
need to modify the equation (\ref{DP}) and (\ref{helop}). The boundary
conditions which ensure vanishing of the boundary variation (\ref{Bbouvar})
now read (after passing to an associated linear problem -- see
a remark between eqs. (\ref{isbc}) and (\ref{sbc}) above):
\begin{equation}
(\Pi X)|_{\partial M}=0,\qquad 
(1-\Pi)(\partial_\sigma +iB\partial_\tau ) 
(1-\Pi )X |_{\partial M}=0 \,.\label{Bsbc}
\end{equation}
Note, that due to the presence of the imaginary unit in front
of $B$ in the boundary conditions (\ref{Bsbc}) neither $P$ nor
$\Delta$ is self-adjoint. This happens even if the projector
$\Pi$ is local as for the $D$-branes. There are two ways to overcome
this difficulty. One can either change rules of continuation to
the Euclidean space for the $B$ field \cite{Kummer:2000ae}, or
assume more complicated conjugation rules in the Euclidean space
\cite{Osborn:1991gm}. We take the second option here to be close
to commonly accepted notations.

It is interesting to note that only a part of the components
of $B$ enter the boundary conditions (\ref{Bsbc}). Let us introduce
the $\gamma_5$ matrix as $\gamma_5=\gamma_\sigma \gamma^\tau$,
$\gamma_5^2=-1$. We split $B$ into odd end even parity parts
with respect to this $\gamma_5$:
\begin{equation}
B=B^{(+)}+B^{(-)},\quad B^{(\pm)}=\frac 12 (B\pm\gamma_5 B\gamma_5),
\quad \gamma_5 B^{(\pm)}=\mp B\gamma_5 \,.\label{oddeven}
\end{equation} 
$B^{(-)}$ commutes with $\Pi_-$, $\Pi_+$ and $\Pi_0$.
$B^{(+)}$ intertwines $\Pi_+$ and $\Pi_-$: $B^{(+)}\Pi_+=\Pi_-B^{(+)}$.
We see, that in the boundary conditions (\ref{Bsbc}) the
components $B^{(+)}$ can interact with $\Pi_0X$ only (if $\Pi_0$ is
included in $1-\Pi$), i.e. with zero modes of $\partial_\tau$.
However, if $\partial_\tau X|_{\partial M}=0$ the term with $B$
in the boundary conditions (\ref{Bsbc}) vanishes. We conclude,
that $B^{(+)}$ drops out completely. $B^{(-)}$ plays a role
of a projection of $B$ on the brane directions. An unexpected feature is
that this projection is {\it local} in the target space, as
in the $D$-brane case, even though the boundary conditions
are non-local. In the rest of the paper we suppose $B^{(+)}=0$
and, therefore, $B=B^{(-)}$.

To discuss the non-commutativity issues we need a convenient representation
for the propagator. Let us put temporarily our system in a box in the
$\tau$-direction so that spectrum of $\partial_\tau$ becomes discrete.
Then 
\begin{eqnarray}
\Pi_+&=&\frac 1{2\pi i}\lim_{\mu\to +0}\int\limits_{-\infty}^{\infty}
\frac{d\alpha}{\alpha -i\epsilon} \exp (i\alpha ({\cal B}+\mu)) \,,\nonumber \\
\Pi_++\Pi_0&=&\frac 1{2\pi i}\lim_{\mu\to +0}\int\limits_{-\infty}^{\infty}
\frac{d\alpha}{\alpha -i\epsilon} \exp (i\alpha ({\cal B}-\mu))
\label{intrep}
\end{eqnarray}
with the boundary helicity operator ${\cal B}$ (\ref{helop}). $\Pi_-$ can be
constructed in the same way. Now we take the limit of an infinite size
of the box and consider (\ref{intrep}) as definitions of the projectors
for that case. 

We can extend the projectors $\Pi$ and $(1-\Pi )$ inside the manifold
$M$ to a neighbourhood of $\partial M$. For $x$ and $y$ in this
neighbourhood,
the propagator with the boundary conditions (\ref{Bsbc}) is
\begin{equation}
\langle X (x) X(y) \rangle = \Pi_x {\cal G}_D(x,y) \Pi_y
+(1-\Pi )_x {\cal G}_N(x,y) (1-\Pi )_y \,,  
\label{prop}
\end{equation}
where ${\cal G}_D$ and ${\cal G}_N$ are the standard propagators
with Dirichlet and generalized Neumann boundary conditions:
\begin{equation}
{\cal G}_D(x,y)\vert_{x\in \partial M}=0,\quad
(\partial_\sigma^{(x)} +iB\partial_\tau^{(x)} )
{\cal G}_D(x,y)\vert_{x\in \partial M}=0 \,.\label{DNpro}
\end{equation}
If $x$ and $y$ are on the boundary, only the Neumann
part of the propagator \cite{Schomerus:1999ug,Seiberg:1999vs}
\begin{equation}
{\cal G}_N(\tau,\tau')=-D \log (\tau -\tau')^2 +\frac i2 \theta
\epsilon (\tau -\tau') \label{Npro}
\end{equation}
survives. Here $\epsilon$ is the sign function and
\begin{equation}
D=(1-B^2)^{-1},\qquad \theta = -B (1-B^2)^{-1} \label{Dtheta}
\end{equation}
are symmetric and antisymmetric matrices respectively. It can be
demonstrated that the $D$-part of the propagator remains
symmetric in $\tau$ and $\tau'$ after the action of the
projectors $(1-\Pi )$. This part does not contribute
to the time-ordered product. Let $\tau\ne\tau'$.
Then ${\cal B}_\tau \theta \epsilon (\tau -\tau')=0$. Hence 
only the $\mu$-term remains in the exponential
in the integral representation for the projectors (\ref{intrep}). 
This means
\begin{equation}
(1-\Pi )_\tau \theta \epsilon (\tau -\tau') (1-\Pi )_{\tau'}=
\theta \epsilon (\tau -\tau') \label{1616}
\end{equation}
if projector on the zero modes $\Pi_0$ is included in $(1-\Pi )$,
and
\begin{equation}
(1-\Pi )_\tau \theta \epsilon (\tau -\tau') (1-\Pi )_{\tau'}=0
\label{1617}
\end{equation}
otherwise. Consequently, we reproduce the famous relations for
a non-commutative space \cite{Schomerus:1999ug,Seiberg:1999vs}
\begin{equation}
[X(\tau),X(\tau')]=i\theta \label{noncom}
\end{equation}
in the former case ($(1-\Pi)\Pi_0=\Pi_0$) and loose the non-commutative
structure in the latter one ($(1-\Pi)\Pi_0=0$). 

In a non-linear theory (as if $G_{\mu\nu}$ and $B_{\mu\nu}$ are not constants)
one can use the background field formalism and repeat this
construction with the part of the action which is quadratic in
quantum fluctuations. Let us outline briefly how the above procedure
can be generalized. 
First we split $X=\bar X +\xi$ into a background
part $\bar X$ and fluctuations $\xi$. The propagator is defined by
the quadratic part of the action:
\begin{equation}
S_2=\int\limits_M d^2x\sqrt{h} \xi D \xi +\int\limits_{\partial M}
d\tau\xi(-\partial_\sigma +L)\xi \,,\label{backact}
\end{equation}
where $D$ and $L$ are some operators. Next we represent $D$ as
\begin{equation}
D=P^\dag P +c \label{reprD}
\end{equation}
with a Dirac operator $P$ and a constant $c$ and construct the
projector $\Pi$ on the Dirichlet components for the corresponding
elliptic complex of Dirac type (see \cite{Atiyah:1980jh,Grubb95,Dowker:2000sy}
for the details). Boundary conditions read:
\begin{equation}
\Pi \xi\vert_{\partial M}=0,\qquad 
(1-\Pi)(-\partial_\sigma +L)(1-\Pi)\xi\vert_{\partial M}=0\,.
\label{bcb}
\end{equation}
Some comments are in order. In Appendix, by using a kind of the degrees
of freedom counting, we argue
that the representation (\ref{reprD}) is generically
possible at least locally. From that consideration
it is clear that explicit construction the operator $P$ for any given $D$
is extremely complicated. This however seems to be not a very serious
drawback. First, eq. (\ref{reprD}) can be viewed as another parametrisation
of the background fields in the closed string sector. Second, constructions
of this type become much easier in the presence of supersymmetry of
the background. And third, the operator $P$ plays an auxiliary role.
It may be not necessary to have an explicit solution of (\ref{reprD})
everywhere on $M$.

\section{T-duality}
Action of $T$-duality transformations on open string theories
has been studied in 
\cite{Dai:1989ua,Leigh:1989jq,Horava:1989ga,Green:1991et}
and since then became a very popular
subject mostly due to the Polchinski's idea \cite{Polchinski:1995mt}  
that $D$-branes may carry R-R charges.
In this section we follow the approach of \cite{Alvarez:1996up}
based on gauging the space-time isometries of the open
string (see also related works \cite{Dorn:1996an,Forste:1996hy}).
Since in our construction two-dimensional blocks play a special
role, we need two commuting isometries acting along the directions
$X^\alpha =k^\alpha_\mu X^\mu$ ,$\alpha =1,2$. 
$k_\alpha$ are the Killing vectors. We suppose that mixed comonenets
of the target space metric are absent, $G_{\alpha \mu}=0, \mu \ne 1,2$.
We choose representation of the Clifford
algebra in such a way that $X^1$ and $X^2$ belong to one irreducible
piece. With this choice, boundary conditions for $X^\alpha$
become fully decoupled from that for the other components
and can be considered separately. Note, that the boundary conditions
for $X^\mu$, $\mu \ne 1,2$ should not be necessarily spectral.
We can consider a configuration which look as an open
string or as a $D$-brane from that directions.
 
For simplicity we neglect the $B$-field.
The action for open strings with gauged isometries reads:
\begin{equation}
S_{{\mbox {\scriptsize {gauged}}}}=\frac 12 \int\limits_M
d^2x \left( G_{\mu\nu}D^a X^\mu D_a X^\nu +iY_\alpha \epsilon^{ab}
(\partial_a V_b^\alpha -\partial_b V_a^\alpha ) \right) \,,
\label{gauged}
\end{equation}
where $D_a X^\mu =\partial_a X^\mu +k^\mu_\alpha V^\alpha_a$. 
The action (\ref{gauged}) possesses a local symmetry
\begin{equation}
\delta X^\mu =\varepsilon^\alpha k_\alpha^\mu ,\qquad
\delta V_a^\alpha =-\partial_a \varepsilon^\alpha \,.
\label{gsym}
\end{equation}
The auxiliary field
$Y^\alpha$ generates the constraint
\begin{equation}
(\partial_a V_b^\alpha -\partial_b V_a^\alpha )\epsilon^{ab}=0
\label{constr}
\end{equation}
which eliminates the gauge field $V$ and reduces (\ref{gauged})
to the standard action (\ref{opact}). On the other hand,
the symmetry (\ref{gsym}) 
allows to gauge away two components $X^1$ and $X^2$.
After integrating by parts the
action for $V$ becomes purely algebraic. Eliminating
$V$ by means of its' equations of motion 
\begin{equation}
G_{\alpha\beta}V_b^\beta h^{ab} =i\epsilon^{ba} \partial_b Y_\alpha
\label{Veom}
\end{equation} 
one arrives at a
dual action with $Y_\alpha$ instead of $X^{1,2}$:
\begin{equation}
S_{{\mbox {\scriptsize {dual}}}}=\frac 12 \int\limits_M
d^2x \left( G_{jk}\partial^a X^j \partial_a X^k +G^{\alpha\beta}
\partial^a Y_\alpha \partial_a Y_\beta \right)\,, 
\label{Sdual}
\end{equation}
where $j,k=3,\dots,d$. The target space metric is inverted
along direction of the first two coordinates.

In the presence of boundaries we should take care of proper
boundary conditions for all fields. Let the field $X^\alpha$
satisfy Dirichlet boundary conditions. We like to deal with
homogeneous boundary value problem, so that $X^\alpha|_{\partial M}=0$.
These boundary conditions are preserved by the gauge transformations
(\ref{gsym}) if the parameter $\varepsilon$ also vanishes at the
boundary, $\varepsilon|_{\partial M}=0$. From the transformation
low for $V_a$ it follows immediately that $V_\tau|_{\partial M}=0$.
A bit more care is needed to define boundary conditions for $V_\sigma$.
Let us introduce a basis in the functional space of the gauge
parameters $\epsilon$ which (similarly to that for the $X$) consists
of eigenfunctions of the Laplace operator satisfying Dirichlet
boundary conditions: $\Delta \epsilon_\lambda =\lambda \epsilon_\lambda$.
Then
\begin{equation}
\partial_\sigma (\partial_\sigma \epsilon_\lambda )\vert_{\partial M}=
(-\lambda -\partial_\tau^2 ) \epsilon_\lambda \vert_{\partial M}=0\,.
\label{2norm}
\end{equation} 
{}From (\ref{2norm}) and (\ref{gsym}) we conclude that $V_\sigma$
satisfies Neumann boundary condition\footnote{Such mixed 
conditions on a vector field when the normal component
satisfies Neumann and the tangential one - Dirichlet boundary
conditions are called {\it relative} boundary conditions in the mathematical
literature \cite{Gilkey}.}. In a similar way we may show that
the field strength $\partial_aV_b-\partial_bV_a$ satisfies 
Neumann boundary conditions as well. Since the field strength
is subject to boundary conditions, the unrestricted field $Y_\alpha$
contains more components than needed
 to generate the 
delta-function $\delta (\partial_aV_b-\partial_bV_a)$. 
To remove this redundancy one should impose some boundary conditions
on $Y_\alpha$ as well. From the equations of motion (\ref{Veom})
it is clear that these should be Neumann boundary conditions.
This choice also
allows to integrate by part in the second term in (\ref{gauged}).
Of course, starting with Neumann boundary conditions for
$X^\alpha$ one will arrive at Dirichlet boundary condition
for $Y_\alpha$. 

We have reproduced the known fact that $T$-duality interchanges
Dirichlet and Neumann boundary conditions. The main advantage
of our slightly modified standard derivation is that it can be
applied to any eigenmode of the operator ${\cal B}$ (\ref{helop})
separately. The reason is that ${\cal B}$ commutes with all relevant
operators. We conclude that $T$-duality interchanges the two by two
blocks in the projectors
$(\Pi )_{\alpha\beta}$ and $(1-\Pi )_{\alpha\beta}$
in the boundary conditions (\ref{sbc}) and, therefore, maps
an $S$-brane to another $S$-brane. 

If we turn on the $B$-field by gauging the bulk action
(\ref{Bvol}) the equations of motion (\ref{Veom}) will
be modified accordingly. Consequentyl, the (dualized)
$B$-field will enter Neumann boundary conditions in the
usual way. Now, $X^\alpha$ or $Y_\alpha$ can exhibit a
non-commuative structure depending on whether the zero
modes are included in corresponding projectors on the
Neumann sub-spaces (see discussion in the previous section).
Since the duality transformation interchanges Dirichlet and
Neumann modes, only the coordinates $X^\alpha$ or their
duals $Y_\alpha$ can be non-commutative (but not $X^\alpha$
{\it and} $Y_\alpha$ simultaneously). This goes in parallel
with the ordinary $D$-brane -- open string duality.

The last question which will be discussed in this section is
a possibility to define duality transformation connecting 
$S$-branes with $D$-branes or $S$-branes with open strings.
To be more specific,
let in (\ref{sbc}) $\Pi=\Pi_+$ be a projector on positive boundary
helicity states. The projector $\Pi_+$ can be extended inside the
manifold $M$. Then we may write the action (\ref{gauged}) where
only the positive boundary helicity modes of the
isometries are gauged assuming that $Y=Y^+=\Pi_+Y$ and $V=V_+=\Pi_+V$ 
also have positive
boundary helicity. Since our $T$-duality arguments may be applied
mode-by-mode, this model is equivalent to the original $S$-brane.
We may eliminate the modes $X_+$ and $V_+$ to obtain the dual
action:
\begin{equation}
S_{{\mbox {\scriptsize {dual}}}}=\frac 12 \int\limits_M
d^2x \left( G_{jk}\partial^a X^j \partial_a X^k +G^{\alpha\beta}
\partial^a Y_\alpha^+ \partial_a Y_\beta^+
+G_{\alpha\beta}\partial^a X^\alpha_- \partial_a X^\beta_- \right)\,, 
\label{newdual}
\end{equation}
where the part of the action containing the non-positive modes $X_-$
remains intact. All $\alpha$ components of the string coordinates
now satisfy Neumann boundary condition, as in the open string theory.
The only difference to the standard case is that non-negative
boundary helicity states interact with the target space metric
$G_{\alpha\beta}$, while the positive helicity states - with the
inverse one $G^{\alpha\beta}$. Therefore, the interaction is
non-local. We managed to trade a non-locality in the boundary
conditions to a non-locality in the bulk action. Construction of
the $S$-brane -- $D$-brane duality goes the same way. Note,
that if the bulk fields are self-dual under the Buscher transformations
\cite{Buscher:1987sk}
(which amounts to $G^{\alpha\beta}=G_{\alpha\beta}$ in the present
simplified case) the action (\ref{newdual}) is local again and
coincides with the original one (\ref{opact}). Boundary conditions
seem to play no role for such configurations.

\section{Conclusions}
In this paper we have introduced a new object, which we
call the $S$-brane, by imposing the spectral boundary conditions
on the coordinates of the endpoints of the bosonic string.
For the simplest geometry of the target space we were able
to study this object quite in detail. This construction
is self-consistent in the sense that it leads to a hermitian
kinetic operator and, hence, to unitary evolution. Depending on
the properties of the non-local boundary projector, the $S$-branes
may be commutative or non-commutative. We have found that only a part
of the components of the $B$-field enters the boundary conditions.
The projection on relevant $B$-field components is done by
a local projector containing the chirality matrix $\gamma_5$.
This indicates that the low energy limit will probably
contain chiral boundary fields.

For the target spaces admitting two commuting isometries
we constructed a $T$-duality transformation of the $S$-brane.
This transformation maps $S$-brane to another $S$-brane
of the same dimension.
This is a rather unusual property for the open string
theory. It suggests that $S$-branes may play a special
role in string dualities. We have also considered $S$-brane --
$D$-brane and $S$-brane -- open strings dualities. By that
transformations we were able to shift non-locality from the
boundary conditions to the bulk action.

Of course, many question regarding the status of these new
objects still remain to be answered. Intrinsic non-locality
of the $S$-branes indicates that they are certain composite
objects. Therefore, it would be an important development
to obtain an $S$-brane as a solitonic solution in another
string model. Another interesting topic is inclusion of the
supersymmetry. We cannot say much on this at the moment, 
except that the spectral boundary conditions by their
construction should be rather friendly to supersymmetry.

A more straightforward development of the formalism presented
in this paper would consist in inclusion of more general
background fields. To this end one should refine the
arguments on the correspondence between Laplace and Dirac
operators (see Appendix). The Dirac parametrisation can be
considered as an alternative way to introduce background
fields in the closed string sector. Exact relations between
the two sets of the background fields are non-local and
rather complicated. In the view of our duality arguments
this is not an unexpected feature. $S$-branes can be obtained
by dualization of $D$-branes or open strings at the expense
of introducing non-localities in the bulk action. 

As a part of the program related to general interaction with
the background one should calculate the $\beta$-functions
for the spectral boundary conditions and reproduce, as a first
step, the Born-Infeld action for the (projected) $B$-field.
This requires development of the heat kernel technique
for these boundary conditions. Note, that the boundary
conditions (\ref{Bsbc}) lead to perhaps the most  complicated
boundary value problem appeared so far in physical applications.
They contain both non-localities and tangential derivatives.
Therefore, calculation of the heat kernel coefficients is not
going to be easy, but is definitely doable. To define the
$T$-duality transformations at the presence of the dilaton
one will also need to calculate the dilaton shift (see
\cite{Schwarz:1993te}). If the heat kernel is known, this can be
done by a rather general method of \cite{Vassilevich:2001kt}.

To conclude, we note  that the $S$-branes seem to be at least
a good testing ground for various string ideas and 
methods.

\section*{Acknowledgments}
The author is grateful to M. Bordag, K. Kirsten,
W. Kummer, E. M. Santangelo and especially to P. Gilkey for 
discussions and/or comments. This work has been supported
by the DFG project BO 1112/11-1.

\section*{Appendix: Dirac and Laplace operators}
In this Appendix we argue that on a two dimensional
Riemannian
manifold $M$ it is generically possible to represent a
Laplace operator via a Dirac operator. We start with
some definitions
following \cite{Dowker:2000sy}. Let ${\cal V}_1$ and ${\cal V}_2$ be
two unitary bundles over $M$ and let $P$ be a first
order partial differential operator 
$P:\ C^\infty ({\cal V}_1)\to C^\infty ({\cal V}_2)$.
Let $P^\dag$ be the formal adjoint of $P$.
We say that $P$ defines an elliptic complex of Dirac type if
the associated second order operators 
\begin{equation}
D_1=P^\dag P \,,\qquad D_2=PP^\dag \label{defDD}
\end{equation}
on $C^\infty ({\cal V}_1)$ and on $C^\infty ({\cal V}_2)$
are of Laplace type - i.e. if these operators have scalar
leading symbol given by the metric tensor on $M$. If 
${\cal V}_1={\cal V}_2$ and $P^\dag =P$, than $P$ is said to be
an operator of Dirac type. We impose a somewhat weaker restriction
on $P$. We identify locally ${\cal V}_1$ and ${\cal V}_2$.
In a local basis $P=\gamma^a\partial_a +r$. We suppose
that $(\gamma^a)^\dag =-\gamma^a$. Then $\gamma$ define
a Clifford module structure. 

We see that by definition any Dirac operator (or complex)
is associated with a formally self-adjoint operator of
Laplace type. Now we wish to invert this construction.
For a given a hermitian operator $D$ of Laplace type we like to find
an elliptic complex of Dirac type such that
\begin{equation}
D=P^\dag P +c\,,\label{repD}
\end{equation}
where $c$ is a constant. We suppose that ${\rm dim}\,{\cal V}$
(the target space dimension) is even so that there exists
a (reducible) representation of the Clifford algebra of
this dimension. There are some global obstructions
to this construction. The simplest one is that $P^\dag P$
is non-negative, while $D$ can have a finite number of
negative modes. One can overcome this obstruction by
choosing $c$ to be the lowest eigenvalue of $D$.
In what follows we will not discuss more global properties of
the decomposition (\ref{repD}). Our analysis will be purely
local.

It can be shown \cite{Gilkey} that any operator
of Laplace type can be represented as
\begin{equation}
E=-(\nabla^a \nabla_a +E)\,,\qquad \nabla_a =\partial_a +\omega_a\,,
\label{anyD}
\end{equation}
with an appropriate connection one-form $\omega$ and an endomorphism 
(a matrix valued function) $E$. It has been demonstrated \cite{BG92}
that we may decompose 
\begin{equation}
P=\gamma^a\tilde\nabla_a +\psi \,,\label{anyP}
\end{equation}
where $\psi$ is a matrix valued function and $\tilde\nabla$
is a compatible unitary connection. That means
$\tilde\nabla_a\gamma^b=0$ and the connection
one-form $\tilde\omega_a$ is anti-hermitian in a suitable basis. 
Even though there
is some arbitrariness in $\tilde\omega$ we suppose that
$\tilde\omega$ is fixed as soon a representation of the
Clifford algebra is fixed. We write
\begin{equation}
P^\dag P=-\tilde\nabla^a\tilde\nabla_a +\frac 14 [\gamma^a,\gamma^b]
\tilde\Omega_{ab}+\gamma^a\tilde\nabla_a\psi 
+\psi^\dag\gamma^a\tilde\nabla_a +\psi^\dag\psi \label{PP}
\end{equation}
with the field strength $\tilde\Omega_{ab}=[\tilde\nabla_a,\tilde\nabla_b]$
including also the Riemann tensor on $M$ so that
the second term in (\ref{PP}) contains the usual $R/4$ contribution. 
Next we compare (\ref{PP}) with (\ref{repD}) and (\ref{anyD})
to obtain
\begin{eqnarray}
&&\omega_a=\tilde\omega_a -\frac 12 (\gamma_a\psi +\psi^\dag\gamma_a) \,,
\nonumber \\
&&E=-\frac 14 [\gamma^a,\gamma^b]\tilde\Omega_{ab}-\psi^\dag\psi
+\frac 12 ((\tilde\nabla^a\psi^\dag )\gamma_a -\gamma_a (\tilde\nabla_a\psi))
\nonumber \\
&&\qquad\qquad\qquad\qquad \qquad\qquad
-\frac 14 (\gamma_a\psi +\psi^\dag\gamma_a)^2
-c \,.\label{psiE}
\end{eqnarray}

Our aim now is to obtain given $\omega$ and $E$ by adjusting
$\psi$ and representation of the Clifford algebra. We are not able
to present a closed form solution. We will rather argue that such
a solution generically exists by using degrees of freedom counting
arguments. 

One can pass from one representation of the Clifford algebra to
another by a local unitary transformation 
$\gamma^a\to U^\dag (x)\gamma^a U(x)$, $U^\dag =U^{-1}$.
By transforming also $\tilde\omega$ and $\psi$ we can force
the operators $P$ and $P^\dag$ to transform homogeneously:
$P\to U^\dag PU$, $P^\dag P\to U^\dag P^\dag P U$.
To analyse solutions on the equation (\ref{repD}) we may fix
a representation of the Clifford algebra and use the corresponding
freedom to gauge transform the operator $D$:
$D\to UDU^\dag$. Therefore, we can
gauge fix $\omega_1$ to coincide with the right hand side of
second of the two equations (\ref{psiE}) containing yet unspecified
function $\psi$. If $d$ is the target space dimension, the 
matrix $\psi$ contains $d^2$ independent components.
Since the operator $D$ is hermitian, the connection $\omega_a$
is typically antihermitian and therefore contains $d(d-1)/2$
independent components. The matrix valued field $E(x)$ should be
hermitian, and thus it contains $d(d+1)/2$ independent entries.
Total number of the degrees of freedom needed to represent
$\omega_2$ and $E$ is $d^2$ at each coordinate point. This is
exactly the freedom contained in $\psi$. We conclude that it
should be possible locally to represent an arbitrary Laplace
type operator $D$ through the Dirac one, as in eq. (\ref{repD}).

Our method is not constructive. Actually it may be extremely
hard to solve the equations (\ref{psiE}). Fortunately, as
we argue at the end of sec.~\ref{secbc}, explicit solution may
be not necessary.

\end{document}